\documentclass[aps,prd,twocolumn,floatfix,nofootinbib,superscriptaddress]{revtex4-1}
\usepackage[colorlinks,linkcolor=blue,anchorcolor=blue,citecolor=blue,urlcolor=blue,breaklinks=true]{hyperref}
\usepackage[libertine]{newtxmath}
\usepackage{graphicx}
\usepackage{slashed}
\usepackage{amsfonts}
\usepackage{amsmath}
\usepackage{blindtext}
\usepackage{microtype}

\newcommand{\vect}[1]{\boldsymbol{#1}}
\newcommand{\ph}[1]{\phantom{#1}}
\newcommand{\sh}[1]{\slashed{#1}}

\graphicspath{{.}{figures/}}

\def\hs{\hspace}

\def\no{\nonumber}

\def\lf{\left}
\def\rg{\right}

\begin{document}

\title{Ground state pseudoscalar mesons on the light front: from the light to heavy sector}

\author{Chao Shi}
\email[]{cshi@nuaa.edu.cn}
\affiliation{Department of Nuclear Science and Technology, Nanjing University of Aeronautics and Astronautics, Nanjing 210016, China}

\author{Ming Li}
\affiliation{Department of Nuclear Science and Technology, Nanjing University of Aeronautics and Astronautics, Nanjing 210016, China}

\author{Xurong Chen}
\affiliation{Institute of Modern Physics, Chinese Academy of Sciences, Lanzhou 730000, China}
\affiliation{Guangdong Provincial Key Laboratory of Nuclear Science, Institute of Quantum Matter, South China Normal University, Guangzhou 510006, China}

\author{Wenbao Jia}
\affiliation{Department of Nuclear Science and Technology, Nanjing University of Aeronautics and Astronautics, Nanjing 210016, China}

\begin{abstract}
We extract the leading Fock-state light front wave functions (LF-LFWFs) of both the light and heavy pseudoscalar mesons, e.g., the pion (at masses of 130 MeV, 310 MeV and 690 MeV), $\eta_c$ and $\eta_b$, from their covariant Bethe-Salpeter wave functions within the rainbow-ladder (RL) truncation. It is shown that the LF-LFWFs get narrower in  $x$ (the longitudinal momentum fraction of meson carried by the quark) with the increasing current quark mass, and the leading twist parton distribution amplitudes (PDAs) inherit this feature. Meanwhile, we find in the pion the LF-LFWFs only contribute around 30\% to the total Fock-state normalization, indicating the presence of significant higher Fock-states within. In contrast, in the $\eta_c$ and $\eta_b$  the LF-LFWFs contribute more than $90$\%, suggesting the $Q\bar{Q}$ valence Fock-state truncation as a good approximation for heavy mesons.  We thus study the 3-dimensional parton distributions  of the $\eta_c$ and $\eta_b$ with the unpolarized generalized parton distribution function (GPD) and the transverse momentum dependent parton distribution function (TMD). Through the gravitational form factors in connection with the GPD, the mass radii of the $\eta_c$ and $\eta_b$ in the light-cone frame are determined to be $r_{E,{\rm LC}}^{\eta_c} =0.150$ fm and $r_{E,{\rm LC}}^{\eta_b} =0.089$ fm respectively. 
\end{abstract}
\maketitle

%===============================================================================
%===============================================================================
\section{INTRODUCTION\label{intro}}
Originating in the Higgs mechanism \cite{Englert:1964et,Higgs:1964pj}, huge current quark mass difference resides in the quantum chromodynamics (QCD), which yield a diverse hadronic phenomenon across the light and heavy sectors. In hard hadronic processes, the quarks' parton nature, which is directly associated with their current masses, gets exposed. The parton structure of the hadrons is thus of great interest  by revealing the substructure of hadrons, and meanwhile experimentally accessible through various hard exclusive and/or inclusive  processes. 

Theoretically, the hadrons' parton structure are  formulated in terms of different sorts of parton distributions. The parton distribution amplitude (PDA), for instance, is an important quantity that incorporates the internal nonperturbative dynamics within the QCD bound state. It serves as a soft input for the factorization of various hard exclusive processes, such as deeply virtual meson production \cite{Radyushkin:1997ki,Vanderhaeghen:1999xj}, B meson decay \cite{Beneke:2001ev,Braguta:2009xu,Zhong:2014fma} and exclusive charmonium $J/\psi+\eta_c$ pair production in $e^+ e^-$ annihilation \cite{Bondar:2004sv,Choi:2007ze}. The determination of the PDAs rely heavily on nonpertubative QCD methods. In the light quark sector, phenomenological models and methods, i.e., QCD sum rule \cite{Chernyak:1983ej,Ball:1998je}, light-front holographic QCD \cite{Brodsky:2011yv} and the Dyson-Schwinger/Bethe-Salpeter equations method (DS-BSEs) \cite{Chang:2013nia,Shi:2014uwa,Shi:2015esa} gave their predictions. Meanwhile, the lattice QCD has predicted it first one or two nontrivial moments \cite{Braun:2006dg,Braun:2015axa,RQCD:2019osh}. Recently, with the help of large momentum effective theory (LaMET) \cite{Ji:2020ect}, the lattice QCD is giving much more information by charting the pointwise behavior of the PDAs \cite{Zhang:2017bzy,Zhang:2017zfe}. Notably, agreement between the lattice QCD and  DS-BSEs is found for pion chiral-extrapolated to physical mass \cite{Zhang:2020gaj}. On the other hand, there is no lattice QCD result on PDAs in the heavy sector yet. Recently, it is proposed that the B meson PDA can be determined from heavy quark effective theory (HQET),  by combining the LaMET and the Euclidean lattice simulation techniques \cite{Wang:2019msf}. As compared to light mesons, heavy mesons are arguably simpler as the quarks move much slower inside, so light front potential models \cite{Choi:2007ze,Li:2017mlw} and nonrelativistic QCD (NRQCD) are applicable \cite{Jia:2015pxx,Wang:2017bgv}.  Meanwhile, QCD sum rule and DSEs also extend from the light sector to the heavy sector and give their predictions on heavy meson DAs \cite{Braguta:2006wr,Zhong:2014fma,Ding:2015rkn,Binosi:2018rht,Serna:2020txe}.

On the other hand, the parton distribution functions (PDF) of hadrons, and in particular their 3-dimensional extension as the generalized parton distribution functions (GPDs)~\cite{Mueller:1998fv,Ji:1996nm,Radyushkin:1997ki} and transverse momentum dependent parton distributions functions (TMDs)~\cite{Collins:2003fm} draw much attention in recent years.  The GPDs provide a unified description of the parton distribution in the longitudinal momentum and transverse spatial coordinates ~\cite{Burkardt:2000za,Burkardt:2002hr}, while the TMDs incorporate the transverse motion of partons and their spin-orbit correlations \cite{Collins:1981uw,Collins:2011zzd}. Meanwhile, the GPDs are connected with hadron matrix elements of the energy-momentum tensor through $x$-weighted moments, which provide valuable information of the spin, energy, and pressure distributions within hadrons~\cite{Ji:1996nm,Burkert:2018bqq,Kharzeev:2021qkd,Wang:2021dis,Wang:2021ujy}. The GPDs and TMDs thus provide much more abundant information on hadron's parton structure. Although present focus on hadrons lies mostly in the light sector, e.g., the nucleon and pion/kaon mesons that are relatively stable \cite{Accardi:2012qut,Chen:2020ijn,Anderle:2021wcy}, it is of theoretical interest to look into the 3D structure of the heavy hadrons with the help of GPD and TMD.

As light-cone quantities, the PDA, GPD and TMD are interconnected by the light-front wave functions (or light-cone wave functions).  The PDA is the LF-LFWF integrated over transverse momentum $\vect{k}_T$ \cite{Lepage:1980fj}, and the GPD and TMD can be calculated with overlap representations in terms of LFWFs \cite{Diehl:2003ny,Diehl:2000xz,Ji:2002xn,Pasquini:2014ppa}. The standard way to obtain the LFWFs is by diagonalizing the light-cone Hamiltonian. Challenges lie in the construction of light-cone Hamiltonian in connection with QCD, as well as its diagonolization when more Fock-states are involved \cite{Brodsky:1997de}. Recently, with the help of basis light-front quantization (BLFQ) technique, the $|q\bar{q}g \rangle$ state is included  in pion and the calculation thus goes beyond the leading Fock-state truncation \cite{Lan:2021wok}. On the other hand, an alternative approach exists by extracting the LFWFs from hadrons' covariant wave functions in the ordinary space-time frame, namely the instant form \cite{tHooft:1974pnl,Liu:1992dg,Heinzl:2000ht,Burkardt:2002uc,Ji:2002xn}. Using this approach, we obtained the LF-LFWFs of the pion and kaon \cite{Shi:2018zqd,Shi:2020pqe}, and later the vector mesons $\rho$ and $J/\psi$ \cite{Shi:2021taf}. A unique advantage of this approach is that it circumvents the light-cone Hamiltonian construction and diagonalization, and allows the extraction of LF-LFWFs from many Fock-states embedded.  In this work, based on the study of the pion at physical mass in \cite{Shi:2020pqe}, we will predict LF-LFWFs of fictitious  pion at masses of 310 MeV and 690 MeV which are directly accessible by lattice QCD, as well as $\eta_c$ and $\eta_b$ from the heavy sector. 

The pseudoscalar meson sits in a special position as in the chiral limit it is the Goldstone boson of dynamical chiral symmetry breaking (DCSB). The pion is thus dominated by the DCSB phenomenon while the Higgs mechanism is almost irrelevant \cite{Roberts:2021xnz}. However, in the heavy mesons, the situation is the opposite: the Higgs mechanism generates most of the quark masses (and consequently the hadron mass) but the DCSB effect weakens. The pseudoscalar mesons across the light and heavy sectors thus provide a good window to observe how the LF-LFWFs evolve with the strength shift between the DCSB and Higgs mechanism. The present study is therefore motivated in several directions: In the light quark sector, we study the LF-LFWFs of the pion at the physical mass $(\approx 130 {\rm MeV})$ and make predictions at testing masses of $m_\pi=310$ MeV and $690$ MeV. The later two cases are chosen as they are directly accessible in lattice simulation \cite{Zhang:2020gaj}. Note that lately the authors of \cite{Ji:2021znw} have proposed a way to extract the hadron LFWFs through lattice QCD based on LaMET, so lattice results can be expected. In the heavy sector, we report the $\eta_c$ and $\eta_b$ LF-LFWFs determined for the first time from the DS-BSEs. Using these LFWFs, we analyze the PDAs of  $\eta_c$ and $\eta_b$ and investigate their 3D parton structure with the help of GPD and TMD.

This paper is organized as follows: In Sec.~\ref{sec:LF-LFWF} we introduce the DS-BSEs formalism and extract from the covariant BS wave functions the LF-LFWFs. In Sec.~\ref{sec:LFWF}, we show the calculated LF-LFWFs of pion (at different masses), $\eta_c$ and $\eta_b$, as well as their PDAs. A comparison is made between the light and heavy mesons. In Sec.~\ref{sec:3d}, the parton structure of $\eta_c$ and $\eta_b$ are studied by means of the GPD (at zero skewness) and unpolarized TMD. Finally we conclude in Sec.~\ref{sec:con}

\section{From Bethe-Salpeter wave functions to LF-LFWFs\label{sec:LF-LFWF}}

Within the DS-BSEs framework, the mesons are treated as bound states and described by their covariant BS wave functions. Within the rainbow-ladder (RL) truncation, which is taken throughout this work, the pseudoscalar mesons can be solved by aligning the quark's DSE for full quark propagator $S(k)$ and meson's BSE  for BS amplitude $\Gamma_M(k,P)$ \cite{Maris:1997tm}, i.e.,
\begin{align}
S(k)^{-1} &= Z_2 \,(i\gamma\cdot k + Z_4 m(\mu)) \nonumber \\ 
 \hspace{10mm} &+ Z_2^2 \int^\Lambda_\ell\!\! {\cal G}(\ell)
\ell^2 D_{\mu\nu}^{\rm free}(\ell)
\frac{\lambda^a}{2}\gamma_\mu S(k-\ell) \frac{\lambda^a}{2}\gamma_\nu
\label{eq:DSE}
\end{align}

\begin{align}
\Gamma_M(k;P) &=-Z_2^2\int_q^\Lambda\!\!
{\cal G}((k-q)^2)\, (k-q)^2 \, \nonumber \\
&\times D_{\mu\nu}^{\rm free}(k-q)
\frac{\lambda^a}{2}\gamma_\mu S(q_+)\Gamma_M(q;P) S(q_-) \frac{\lambda^a}{2}\gamma_\nu ,
\label{eq:BSE}
\end{align}
Here $\int^\Lambda_q$ implements a Poincar\'e invariant regularization of the four-dimensional integral, with $\Lambda$ the regularization mass-scale. The $D^{\rm{free}}_{\mu\nu}$ is the free gluon propagator in the Landau gauge. The quark momentum partition $q_{\pm}=q\pm P/2$. The $m(\mu)$ is the current-quark mass renormalized at scale of $\mu$. The $Z_{2}$ and $Z_4$ are the quark wave function and mass renormalization constants respectively. Here a factor of $1/Z_2^2$ is picked out to preserve multiplicative renormalizability in solutions of the DSE and BSE \cite{Bloch:2002eq}. The Bethe-Salpeter amplitudes are eventually normalized canonically
\begin{align}
2 P_\mu&=\int_k^\Lambda \left\{{\rm Tr}\left[\bar{\Gamma}_M(k;-P)\frac{\partial S(k_+)}{\partial P_\mu} \Gamma_M(k;P)S(k_-)\right] \right. \nonumber \\
&\hspace{10mm}+\left. {\rm Tr}\left[\bar{\Gamma}_M(k;-P)S(k_+)\Gamma_M(k;P)\frac{\partial S(k_-)}{\partial P_\mu} \right]\right \},
\end{align}
with $\bar{\Gamma}_M(k,-P)^T=C^{-1}\Gamma_M(-k,-P)C$ and $C=\gamma_2 \gamma_4$. 
Notably, the RL truncation preserves the (near) chiral symmetry of QCD by respecting the axial vector Ward-Takahashi identity \cite{Maris:1997hd}. The DCSB is therefore faithfully reflected and the Goldstone nature of light pseudoscalar mesons are manifested. Meanwhile, the RL truncation also applies to the heavy mesons \cite{Blank:2011ha,Hilger:2014nma,Fischer:2014cfa}. We therefore solve the pion and heavier $\eta_c$ and $\eta_b$ mesons in the same RL truncation.

The modeling function ${\cal G}(l^2)$ in Eqs.~(\ref{eq:DSE},\ref{eq:BSE}) absorbs the strong coupling constant $\alpha_s$, as well as the dressing effect in both the quark-gluon vertex and full gluon propagator. Popular models include the Maris-Tandy (MT) model, and the later Qin-Chang (QC) model
\cite{Qin:2011xq}
\begin{equation}
\label{eq:GQC}
{\cal G}(s) = \frac{8 \pi^2}{\omega^4} D  \, {\rm e}^{-s/\omega^2}
+ \frac{8 \pi^2 \gamma_m}{\ln [ \tau + (1+s/\Lambda_{\rm QCD}^2)^2]} {\cal F}(s).
\end{equation} 
The second term in Eq.~{\ref{eq:GQC}} is the perturbative QCD result \cite{Maris:1997tm,Qin:2011xq} that describes the UV behavior, while the first term incorporates essentially nonperturbative dressing effects at low and moderate momentum. As compared to the MT model, the QC model improves the far infrared behavior of gluon propagator to be in line with lattice QCD \cite{Cucchieri:2008fc,Bogolubsky:2009dc} and modern DSE  study \cite{Aguilar:2008xm}, while in hadron study the two are equally good.  Historically, combined with the RL truncation, the MT and/or QC models  well describe a range of hadron properties, including the pseudoscalar and vector meson masses, decay constants and various elastic and transition form factors \cite{Maris:1997hd,Maris:1999nt,Maris:2000sk,Jarecke:2002xd,Bhagwat:2006pu,Xu:2019ilh}.

The model parameters in this work are set up as follows. For pion, we consider three cases, i.e., $m_\pi = 130$ MeV, $m_\pi= 310$ MeV and $m_\pi= 690$ MeV. The first one is physical while the later two are for exploratory purpose, but directly accessible in lattice QCD simulations. In Eq.~(\ref{eq:GQC}), we employ the well determined parameters $\omega=0.5$ GeV, $D=(0.82 {\rm GeV})^3/\omega$ \cite{Qin:2011xq}. Meanwhile, we omit the UV term of Eq.~(\ref{eq:GQC}), which determines the UV behavior of pion's BS amplitude. Physically, this means we will only focus on the low and moderate $\vect{k}_T$ part of the pion's LF-LFWFs, but discard the their UV part. In this case, Eqs.~(\ref{eq:DSE},\ref{eq:BSE}) are super-renormalizable and the renormalization constants $Z_2$ and $Z_4$ can be set to 1. The only remaining parameter is the current quark mass and we take $m_{u/d}=5$ MeV, $27$ MeV and 119 MeV, which produce $m_\pi=130$ MeV, 310 MeV and 690 MeV respectively. Note that $m_{u/d}=119$ MeV already reaches the strange quark mass.

In the heavy sector, we take $\omega=0.7$ GeV, $D=(0.765  {\rm GeV})^3/\omega$. They reproduce the physical mass spectrum and decay constants of heavy mesons as $\eta_c$, $J/\psi$, $\eta_b$ and $\Upsilon$. Their deviation from those in the light sector is the consequence of the diminishing of dressing effect from light antiquark-quark-gluon vertex $\bar{q}qg$ to heavy $\bar{Q}Qg$ vertex \cite{Ding:2015rkn}. For the mass parameters, we determine $M_c(m_c^2)=m_c=1.32$ GeV and $M_b(m_b^2)=m_b=4.30$ GeV. Here the $M_{c/b}(k^2)$ are the mass functions of quark propagator defined in Eq.~(\ref{eq:S}). These parameters produce $m_{\eta_c}=2.92$ GeV, $m_{\eta_b}=9.4$ GeV, and decay constants $f_{\eta_c}=0.272$ GeV and $f_{\eta_b}=0.476$ GeV. For comparison, we remind that the particle data group (PDG) gives $m_c = 1.27 \pm 0.02 $ GeV, $m_b=4.18^{+0.03}_{-0.02}$ GeV, $m_{\eta_c}=2.984$ GeV and $m_{\eta_b}=9.398$ GeV \cite{ParticleDataGroup:2020ssz}. For the decay constants, lattice QCD gives  $f_{\eta_c}=0.279$ GeV \cite{Davies:2010ip} and $f_{\eta_b}= 0.472$ GeV \cite{McNeile:2012qf} respectively.

Using these parameters, the $S(k)$ and $\Gamma_M(k;P)$ can be numerically solved with Eqs.~(\ref{eq:DSE},\ref{eq:BSE}).  Note that the $S(k)$ takes the general decomposition 
\begin{align}\label{eq:S}
S(k)=\frac{1}{i A(k^2) \sh{k}+B(k^2)}=\frac{Z(k^2)}{i\sh{k}+M(k^2)}
\end{align}
and $\Gamma_M(k;P)$ takes 
\begin{align}
\label{eq:gammapara}
\Gamma_M(k;P) &= \gamma_5 \Big[i E(k;P)+\sh{P} F(k;P)\nonumber \\
&\hs{10mm}
+(k \cdot P)\sh{k}\,G(k;P)+i [\sh{k},\sh{P}]\,H(k;P)\Big].
\end{align}
The ${\cal F}=E,F,G$ and $H$ are scalar functions of $k^2, k\cdot P$ and $P^2$. In the end, we get the numerical solutions to $A, B$ and ${\cal F}$'s.

To extract the LF-LFWFs, we further parameterize   $S(k)$ and $\Gamma_M(k;P)$ with analytical forms. The $S(k)$ is written as the sum of pairs of complex conjugate poles \cite{Souchlas:2010boa}
\begin{align}
\label{eq:spara}
S(k)=\sum_{i=1}^{N}\left [ \frac{z_i}{i \sh{k}+m_i}+\frac{z^*_i}{i \sh{k}+m^*_i} \right ],
\end{align}
with $N=2$. For the BS amplitude $\Gamma_M(k;P)$, we parameterize its scalar functions ${\cal F=}E, F, G$ and $H$ with Nakanishi-like representation \cite{Nakanishi:1963zz,Shi:2020pqe}
\begin{align}
\label{eq:fpara}
{\cal F}(k;P)&=\int_{-1}^1 d\alpha \rho_{i}(\alpha)\bigg[\frac{U_1 \Lambda^{2 n_1}}{(k^2+\alpha k\cdot P+\Lambda^2)^{n_1}}\nonumber \\ 
&\hs{30mm}
+\frac{U_2 \Lambda^{2 n_2}}{(k^2+\alpha k\cdot P+\Lambda^2)^{n_2}}\bigg]\nonumber\\ 
&+\int_{-1}^1 d\alpha \rho_u(\alpha)\frac{U_3 \Lambda^{2 n_3}}{(k^2+\alpha k\cdot P+\Lambda^2)^{n_3}}, \allowdisplaybreaks[2]\\
\rho_i(\alpha)&=\frac{1}{\sqrt{\pi}}\frac{\Gamma(3/2)}{\Gamma(1)}\Big[C_0^{(1/2)}(\alpha) \no \\
&\hs{22mm}
+ \sigma^i_1 C_1^{(1/2)}(\alpha) +\sigma^i_2 C_2^{(1/2)}(\alpha)\Big],
\label{eq:rho}
\end{align}
where $\rho_u(\alpha)=\frac{3}{4}(1-\alpha^2)$ and  \{$C_n^{(1/2)}, n=0,1,...,\infty$\} are the Gegenbauer polynomials of order $1/2$. The value of the  parameters are listed in Table.~\ref{tab:parapion} and Table.~\ref{tab:paraetaQ}. The outgoing quark and anti-quark in the meson carry momentum $k+P/2$ and $k-P/2$ respectively, so ${\cal F}(k;P)$ is even in $k \cdot P$ due to charge parity.

On the other hand, in the light-cone frame, the pseudoscalar meson $M$ with valence quark $f$ and valence  anti-quark $\bar{h}$ at the leading Fock-state is given by~\cite{Brodsky:1997de,Jia:2018ary}
\begin{align}\label{eq:LFWF1}
|M\rangle &= \sum_{\lambda_1,\lambda_2}\int \frac{d^2 \vect{k}_T}{(2\pi)^3}\,\frac{dx}{2\sqrt{x\bar{x}}}\, \frac{\delta_{ij}}{\sqrt{3}} \nonumber \\
&\hspace{10mm} \Phi_{\lambda_1,\lambda_2}(x,\vect{k}_T)\, b^\dagger_{f,\lambda_1,i}(x,\vect{k}_T)\, d_{h,\lambda_2,j}^\dagger(\bar{x},\bar{\vect{k}}_T)|0\rangle.
\end{align}
where $\vect{k}_T$ is the transverse momentum of the quark $f$, and $x=\frac{k^+}{P^+}$ is the light-cone longitudinal momentum fraction of the active quark. The rest variables are $\bar{x}=1-x$ and  $\bar{\vect{k}}_T=-\vect{k}_T$. The $\lambda_i = (\uparrow,\downarrow)$ denotes the quark helicity and $\delta_{ij}/\sqrt{3}$ is the color factor.  The $b^\dagger$ and $d^\dagger$ are the creation operators for quark and antiquark respectively. The $\Phi_{\lambda_1,\lambda_2}(x,\vect{k}_T)$ are the LFWFs that encode the nonperturbative internal dynamical information.

Meanwhile, in the case of pseudoscalar mesons and abided by the constraint from $\hat{Y}$ parity (The $\hat{Y}$ transform consists a parity operation followed by a 180\textdegree rotation around the $y$ axis \cite{Ji:2003yj}), the four $\Phi_{\lambda_1,\lambda_2}(x,\vect{k}_T)$'s can be expressed with two independent scalar amplitudes \cite{Ji:2003yj}, i.e.,

\begin{align}\label{eq:phiT}
\!\!\!\! \Phi_{\uparrow,\downarrow}(x,\vect{k}_T)&=\psi_0(x,\vect{k}_T^2),&\Phi_{\downarrow,\uparrow}(x,\vect{k}_T)=-\psi_0(x,\vect{k}_T^2), \notag \\
\!\!\!\!  \Phi_{\uparrow,\uparrow}(x,\vect{k}_T)&=k_T^-\psi_1(x,\vect{k}_T^2), &\Phi_{\downarrow,\downarrow}(x,\vect{k}_T)=k_T^+\psi_1(x,\vect{k}_T^2),
\end{align}
where $k_T^{\pm}=k^1\pm ik^2$. The subscripts $0$ and $1$ of $\psi(x,\vect{k}_T^2)$ indicate the orbital angular momentum of the (anti)quark projected onto the light-cone $z-$direction.  The $\psi(x,\vect{k}_T^2)$'s are easier to compute as they contain less variables than $\Phi(x,\vect{k}_T)$. They can be obtained from the Bethe-Salpeter wave function via the light front projections~\cite{Mezrag:2016hnp,Xu:2018eii,Shi:2018zqd}
\begin{align}
\label{eq:psi0}
\psi_0(x,\vect{k}_T^2)&= \ph{-}\sqrt{3}\,i\!\int \frac{dk^+dk^-}{2\,\pi} \nonumber \\
& \hspace{11mm} 
\textrm{Tr}_D\!\left[ \gamma^+ \gamma_5 \chi_{f\bar{h}}(k,P)\right]  \delta\left(x\,P^+ -k^+\right),  \\
\label{eq:psi1}
\psi_1(x,\vect{k}_T^2)&= -\sqrt{3}\,i\!\int \frac{dk^+dk^-}{2\,\pi}\,  \frac{1}{\vect{k}_T^2} \nonumber \\ 
& \hspace{11mm} 
\textrm{Tr}_D\left[ i\sigma_{+ i}\, k_{T}^i\, \gamma_5\, \chi_{f\bar{h}}(k,P) \right] 
 \delta\left(x\,P^+ -k^+\right),   
\end{align}
where the trace is over Dirac indices. The BS wave function can be expressed with quark propagator $S(k)$ and BS amplitude $\Gamma_M(k;P)$ as $\chi_{f\bar{h}}(k;P) = S_f(k+P/2)\,\Gamma_{M}(k;P)\,S_h(k-P/2)$. Using Eqs.(\ref{eq:spara}-\ref{eq:rho}) and Eqs.(\ref{eq:psi0},\ref{eq:psi1}), one can reproduce the pointwise behavior of $\psi(x,\vect{k}_T^2)$ with very high precision, using same method explained in section II of \cite{Shi:2020pqe}.

\section{LF-LFWFs of pseudoscalar mesons}\label{sec:LFWF}

\begin{figure}[tbp]
\centering\includegraphics[width=\columnwidth]{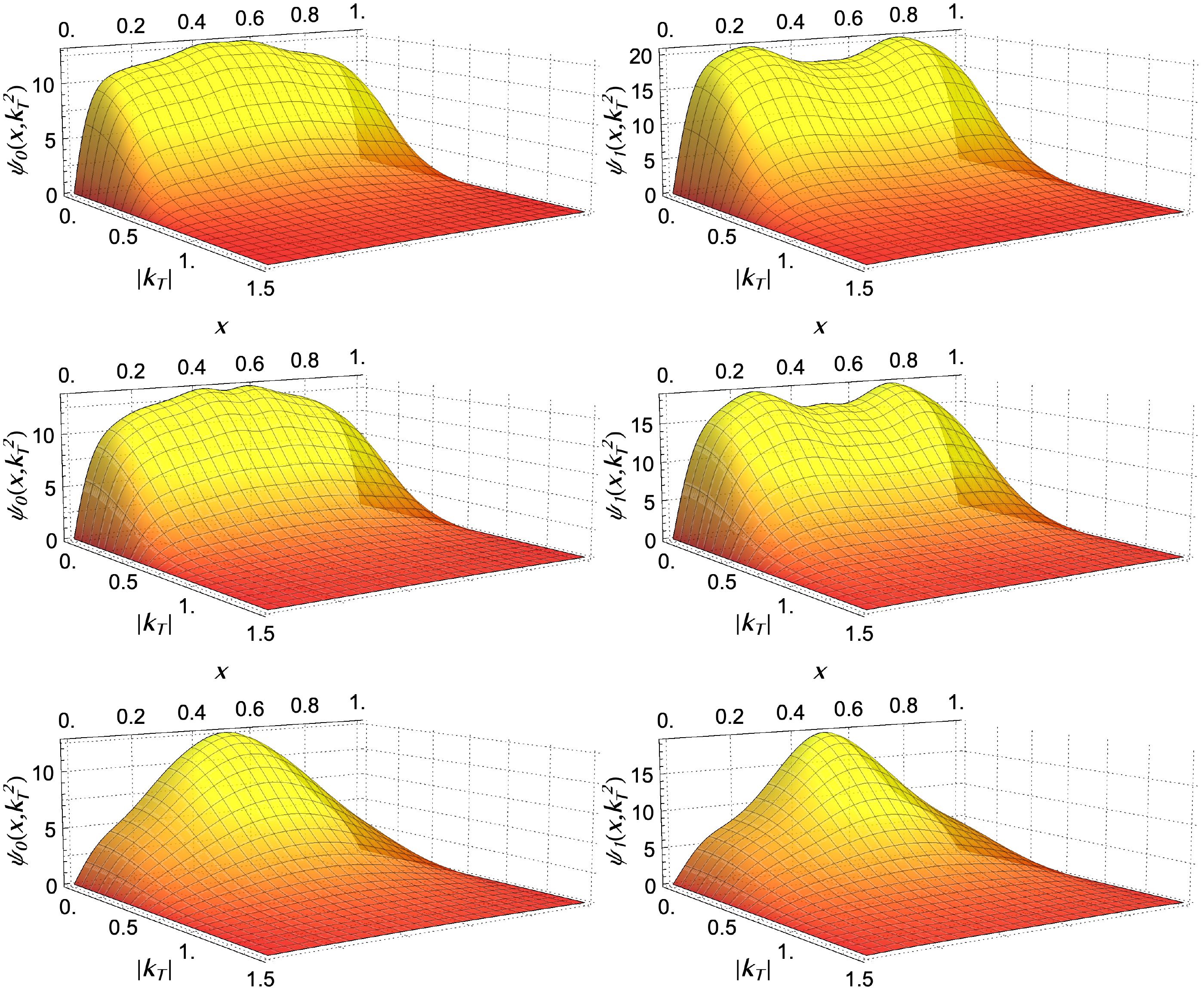}
\caption{The $\psi_0(x,\vect{k}_T^2)$ and $\psi_1(x,\vect{k}_T^2)$ of pion at $m_\pi=130$ MeV (top row), 310 MeV (middle row) and 690 MeV (bottom row). }
\label{fig:psisq}
\end{figure}

\begin{figure}[tbp]
\centering\includegraphics[width=\columnwidth]{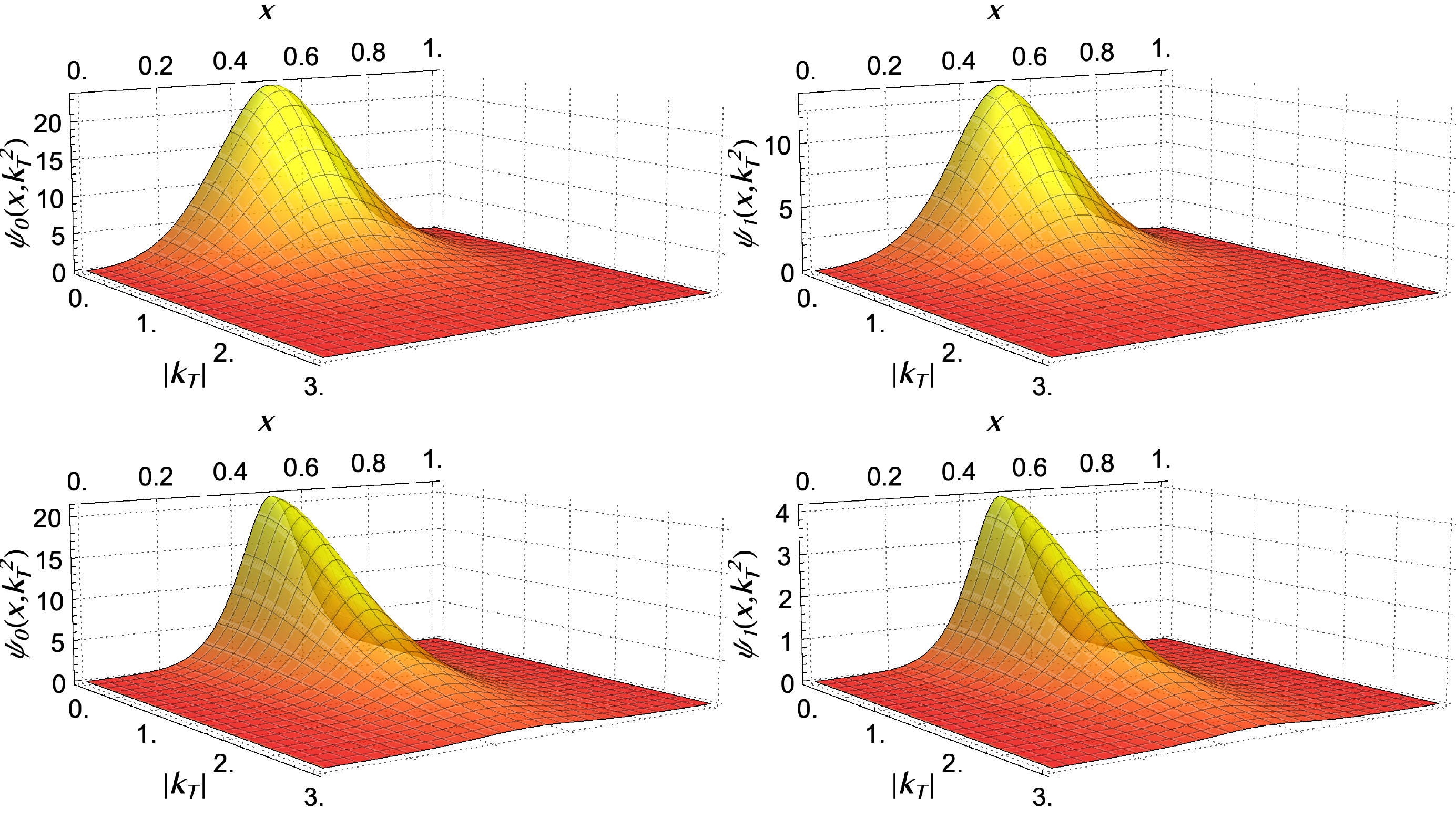}
\caption{The $\psi_0(x,\vect{k}_T^2)$ and $\psi_1(x,\vect{k}_T^2)$ of $\eta_c$ (top row) and $\eta_b$ (bottom row). 
}
\label{fig:psisQ}
\end{figure}

The LF-LFWFs for the pion are shown in Fig.~\ref{fig:psisq}. From the top row to the bottom, we display the results of $m_\pi=130$ MeV, $310$ MeV and $690$ MeV respectively. Comparing these LF-LFWFs, a prominent feature is that as the current quark mass increases, the LF-LFWFs get narrower in $x$. Fig.~\ref{fig:psisq} therefore suggests that the DCSB tends to broaden the $x-$distribution of pseudoscalar mesons LF-LFWFs while the explicit chiral symmetry breaking brought by Higgs mechanism does the opposite. The tendency continues to the heavy sector. In Fig.~\ref{fig:psisQ} we show the LF-LFWFs of $\eta_c$ and $\eta_b$ mesons,  which are significantly narrower than those in the light quark sector. On the other hand, the light meson LF-LFWFs  decrease much faster than the heavy meson in $\vect{k}_T$. This indicates the transverse motion of quarks in heavy mesons are more active than that in the light system.

%===============================================================================
%===============================================================================
%===============================================================================

Aside from the profile, the magnitude of the LF-LFWFs also provide important information of the pseudoscalar mesons' parton structure. Rigorously speaking, the meson's LFWFs of all Fock-states should normalize to unity, i.e., 
\begin{align}
1&=\sum_{\lambda,\lambda'} N_{\lambda,\lambda'} +N_\textrm{HF},
\end{align}
with
\begin{align}
N_{\lambda,\lambda'}&=\int_0^1 dx \int \frac{d \vect{k}_T^2}{2(2 \pi)^3}  |\Phi_{\lambda,\lambda'}(x,\vect{k_T})|^2. \label{eq:N2}
\end{align}
The $N_{\textrm{HF}}$ refers to higher Fock-states contribution. The $N$'s are simply the overlap between the LFWFs and their complex conjugate, so they are positive definite.  Here we infer the value of $N_{\textrm{HF}}$ by subtracting unity with the LF-LFWFs contribution. Our result is listed in Table.~\ref{tab:N}. Apparently, as the system gets heavier, the higher Fock-states contribution diminishes. This suggests i) heavy systems as $c\bar{c}$ and $b\bar{b}$ could be well approximated with their LF-LFWFs and ii) in light system such as pion where the Higgs Mechanism is almost irrelevant, there are lots of higher Fock-states generated in association with the DCSB. We therefore believe this is another novel property of the parton structure of mesons in connection with the DCSB.  

Finally, there is a shift in the relative strength between $\psi_0(x,\vect{k}_T^2)$ and $\psi_1(x,\vect{k}_T^2)$ as the quark mass changes. Defining the ratio
\begin{align}\label{eq:r}
 r \equiv \frac{N_{\uparrow,\uparrow}+N_{\downarrow,\downarrow}}{N_{\uparrow,\downarrow}+N_{\downarrow,\uparrow}}=\frac{N_{\uparrow,\uparrow}}{N_{\uparrow,\downarrow}},
\end{align}
we find $r=0.56, 0.16$ and $0.04$ for pion, $\eta_c$ and $\eta_b$ respectively. Therefore as the spin anti-parallel (S-wave) LFWF $\psi_0(x,\vect{k}_T^2)$ provides more contribution, there is also considerable P-wave component in the pion. In the heavy quarkonium, the S-wave component becomes dominant as the system gets non-relativistic.

\begin{table}[htbp]

\begin{center}
\begin{tabular*}%{llcccccccc}
{\hsize}
{
l@{\extracolsep{0ptplus1fil}}
c@{\extracolsep{0ptplus1fil}}
c@{\extracolsep{0ptplus1fil}}
c@{\extracolsep{6ptplus1fil}}
c@{\extracolsep{0ptplus1fil}}
c@{\extracolsep{0ptplus1fil}}
c@{\extracolsep{0ptplus1fil}}
c@{\extracolsep{0ptplus1fil}}
c@{\extracolsep{0ptplus1fil}}
c@{\extracolsep{0ptplus1fil}}
c@{\extracolsep{0ptplus1fil}}}\hline
   & $N_{\uparrow,\downarrow}=N_{\downarrow,\uparrow}$ & $N_{\uparrow,\uparrow}=N_{\downarrow,\downarrow}$ & $N_{HF}$    & r\\\hline
$\pi$ (130 MeV) & 0.1 & 0.056 & 0.69  &0.56 \\
$\eta_c$ & 0.40 & 0.064 & 0.07 & 0.16 \\
$\eta_b$ & 0.48 & 0.02 & $\approx$ 0.0& 0.04 \\\hline
\end{tabular*}
\end{center}
\vspace*{-4ex}
\caption{LFWFs contribution to Fock-states normalization. See Eq.~(\ref{eq:N2}) for definition of $N$ and Eq.~(\ref{eq:r}) for $r$. 
\label{tab:N}
}
\end{table}

\begin{figure}[htbp]
\centering\includegraphics[width=\columnwidth]{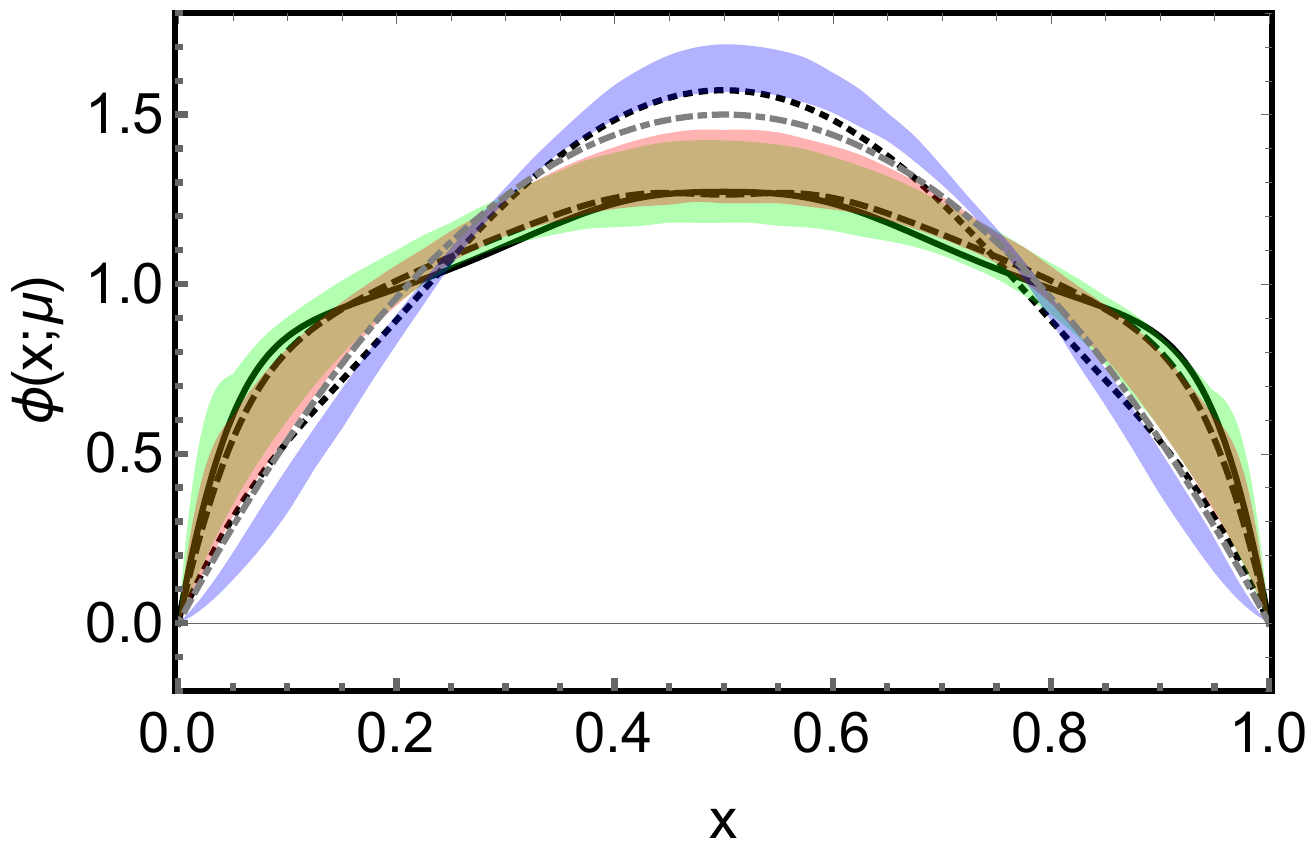} \caption{The PDA of pion at masses of $m_\pi=130$ MeV (solid), 310 MeV (dashed) and 690 MeV (dotted). The colored bands are results from lattice QCD \cite{Zhang:2020gaj} at same masses of $m_\pi=130$ MeV (red), 310 MeV (green) and 690 MeV (blue).}
\label{fig:PDA1}
\end{figure}

We next look at the leading-twist parton distribution amplitudes (PDA) of the mesons. The twist-2 PDA of pseudoscalar meson was originally defined as the $\vect{k}_T$-integrated LFWF \cite{Lepage:1980fj}, i.e.,
\begin{align}
\label{eq:PDAdef}
\phi_\pi(x,Q)&\propto \int_{\vect{k}_T^2 \le Q^2}\frac{d^2 \vect{k}_T}{16 \pi^3}\  \psi_0(x,\vect{k}_T^2),
\end{align}
with the normalization condition
\begin{align}
\int_0^1 dx\phi(x,Q)=1
\end{align}
In Fig.~\ref{fig:PDA1}, we show our results on pion DAs for three masses, i.e., 130 MeV, 310 MeV and 690 MeV, denoted by solid, dashed and dotted curves respectively. The green, red and blue bands are adopted from lattice QCD calculation that employed same pion masses respectively, while at a hadron momentum of $P_z=1.75$ GeV in the LaMET approach \cite{Zhang:2020gaj}. General agreement between the two calculations is found, including the evolution of PDA with increasing current quark mass (or pion mass). For instance, the PDAs of pion with mass 130 MeV and 310 MeV do not differ much, i.e., they are broad and concave functions.  On the other hand, at the mass of 690 MeV, the PDAs get significantly narrower and is close to the asymptotic form $6x(1-x)$ (gray dot-dash-dashed curve), as first pointed out in DS-BSEs \cite{Ding:2015rkn} and confirmed later in \cite{Zhang:2020gaj}. We note that the our PDAs are associated with a scale around 1.0 GeV: since we only took the first term in Eq.~(\ref{eq:GQC}), the model implements a soft momentum cutoff $\approx \Lambda \approx 2 \omega= 1.0$ GeV. Meanwhile the lattice QCD result is at the scale of 2 GeV. Under the Efremov--Radyushkin--Brodsky--Lepage (ERBL) evolution \cite{Efremov:1978rn,Lepage:1980fj}, all curves would evolve very slowly toward the asymptotic form. 

\begin{figure}[htbp]
\centering\includegraphics[width=\columnwidth]{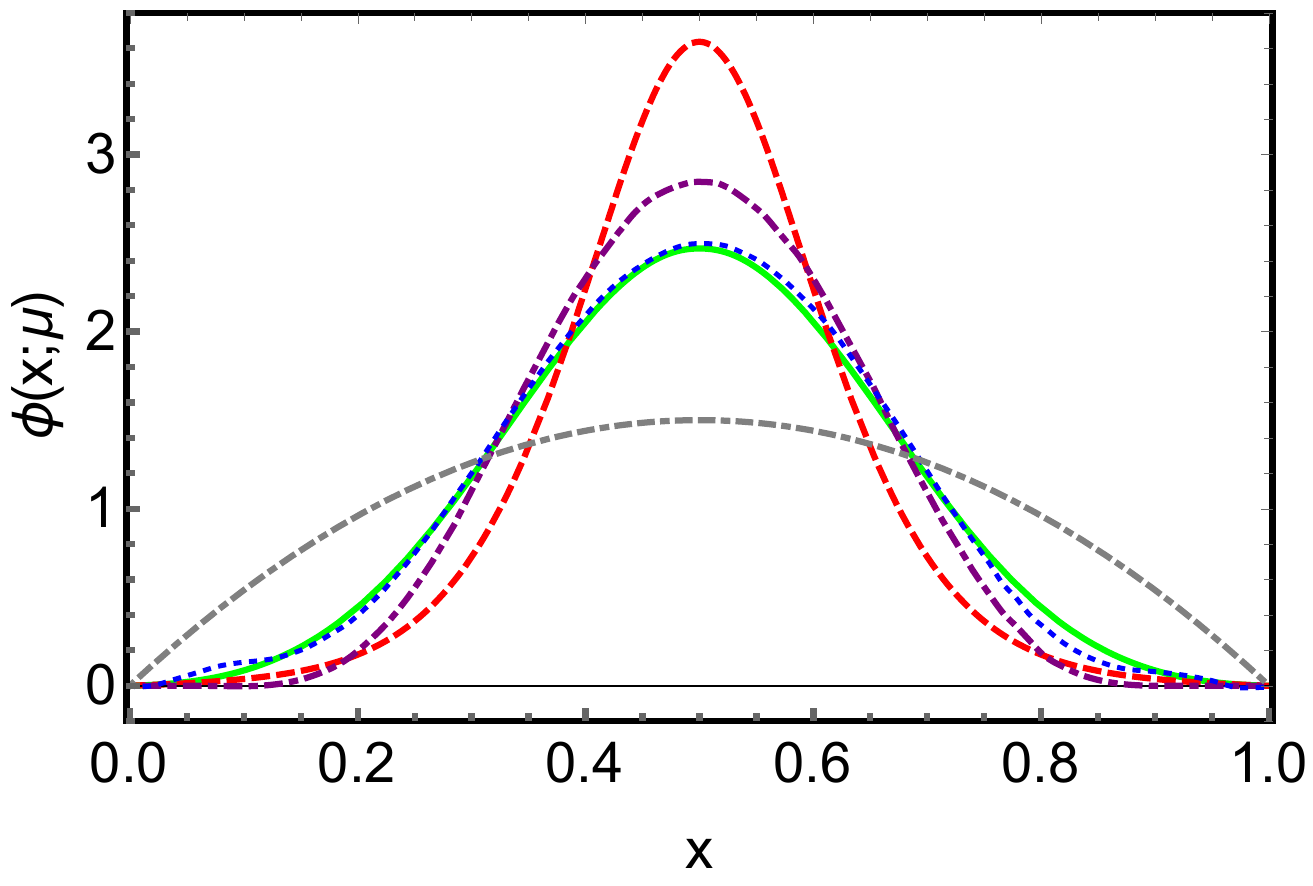} \caption{A comparison of our PDA of $\eta_c$ (green solid) and $\eta_b$ (red dashed) and sum rule results (blue dotted and purple dot-dashed respectively) under the background field theory \cite{Zhong:2014fma}.}
\label{fig:PDA2}
\end{figure}

The PDAs of $\eta_c$ and $\eta_b$ are shown in Fig.~\ref{fig:PDA2}. The green solid curve is our result on $\phi_{\eta_c}(x,Q=m_c)$, with $m_c=1.32 \textrm{GeV}$ in our parameter setup. The dashed red curve is our $\phi_{\eta_b}(x,Q=m_b)$ with  $m_b=4.28 \textrm{GeV}$. For comparison, we show the $\phi_{\eta_c}(x,Q\approx m_c)$ (blue dotted) and $\phi_{\eta_b}(x,Q \approx m_b)$ (purple dot-dashed)  from sum rule calculation \cite{Zhong:2014fma}. One can see the PDAs of $\eta_c$ agree very well. For $\eta_b$, there is some deviation between the two curves. However, the deviation reduces if the uncertainties are considered. For instance, in \cite{Zhong:2014fma} the moments $\langle (2x-1)^m\rangle$ with m=2,4,6 are $0.067 \pm 0.007$, $0.011 \pm 0.002$,$0.003 \pm 0.001$ respectively, while our red dashed curve gives 0.059, 0.0125 and 0.0047. On the other hand, covariant light front model \cite{Hwang:2008qi} and BLFQ approach \cite{Li:2017mlw} all yield semi-quantitatively similar results.  In the absence of lattice QCD calculation, they all suggest that the PDAs of heavy pseudoscalar mesons are narrowly distributed in $x$ as compared to asymptotic form $6x(1-x)$ (gray dot-dash-dashed curve).

\section{Spatial and Momentum Tomography of the heavy pseudoscalar mesons.}\label{sec:3d}

In the light-cone gauge, the unpolarized quark GPD of pseudoscalar meson is defined as
\begin{align}
H^q_M(x,\xi,t;\mu) &= \frac{1}{2}\int\frac{dz^-}{2\pi}\,e^{ixP^+z^-}  \nonumber \\
&\hs*{5mm}
\left< P+\tfrac{\Delta}{2}\left|\bar{\psi}^q(-\tfrac{z^-}{2})\,\gamma^+\,\psi^q(\tfrac{z^-}{2})\right|P-\tfrac{\Delta}{2}\right>. \label{eq:Hdef}
\end{align}
The $x$ is the parton's averaged light-cone momentum fraction and $\xi=-\frac{\Delta^+}{2 P^+}$ is the skewness. The momentum transfer is $t = \Delta^2 = -\frac{4\xi^2 m_M^2+\vect{\Delta}_T^2}{1-\xi^2}$.  The GPD has two distinct domains, where $|x|<|\xi|$ is the ERBL region and $1>|x|>|\xi|$ is the Dokshitzer--Gribov--Lipatov--Altarelli--Parisi (DGLAP) region, named after their evolution with factorization scale $\mu$. Here we will be focusing on the GPD of $\eta_c$ and $\eta_b$ at zero skewness, i.e., the $H^q_M(x,0,t)$. It gives rise to many interesting quantities, e.g., the one-dimensional collinear parton distribution function (PDF), the impact parameter dependent parton distributions (IPDs) and the gravitational form factor (GFF). The overlap representation of $H^f_M(x,0,t)$ in terms of LFWFs reads ~\cite{Diehl:2003ny,Diehl:2000xz,Mezrag:2016hnp,Chouika:2016cmv}
\begin{align}
\label{eq:Hoverlap}
H^q_M(x,0,t;\mu_0)&=\int \frac{d^2\vect{k}_T}{(2\pi)^3}\big[ \psi_0^*(x,\hat{\vect{k}}_T)\,\psi_0(x,\tilde{\vect{k}}_T)\nonumber \\
&\hspace{12mm}
+\hat{\vect{k}}_T \cdot \tilde{\vect{k}}_T\,\psi_1^*(x,\hat{\vect{k}}_T)\,\psi_1(x,\tilde{\vect{k}}_T) \big],
\end{align}
with $\hat{\vect{k}}_T=\vect{k}_T+(1-x)\frac{\vect{\Delta}_T}{2}$ and $\tilde{\vect{k}}_T=\vect{k}_T-(1-x)\frac{\vect{\Delta}_T}{2}$.

\begin{figure}[tbp]
\includegraphics[width=\columnwidth]{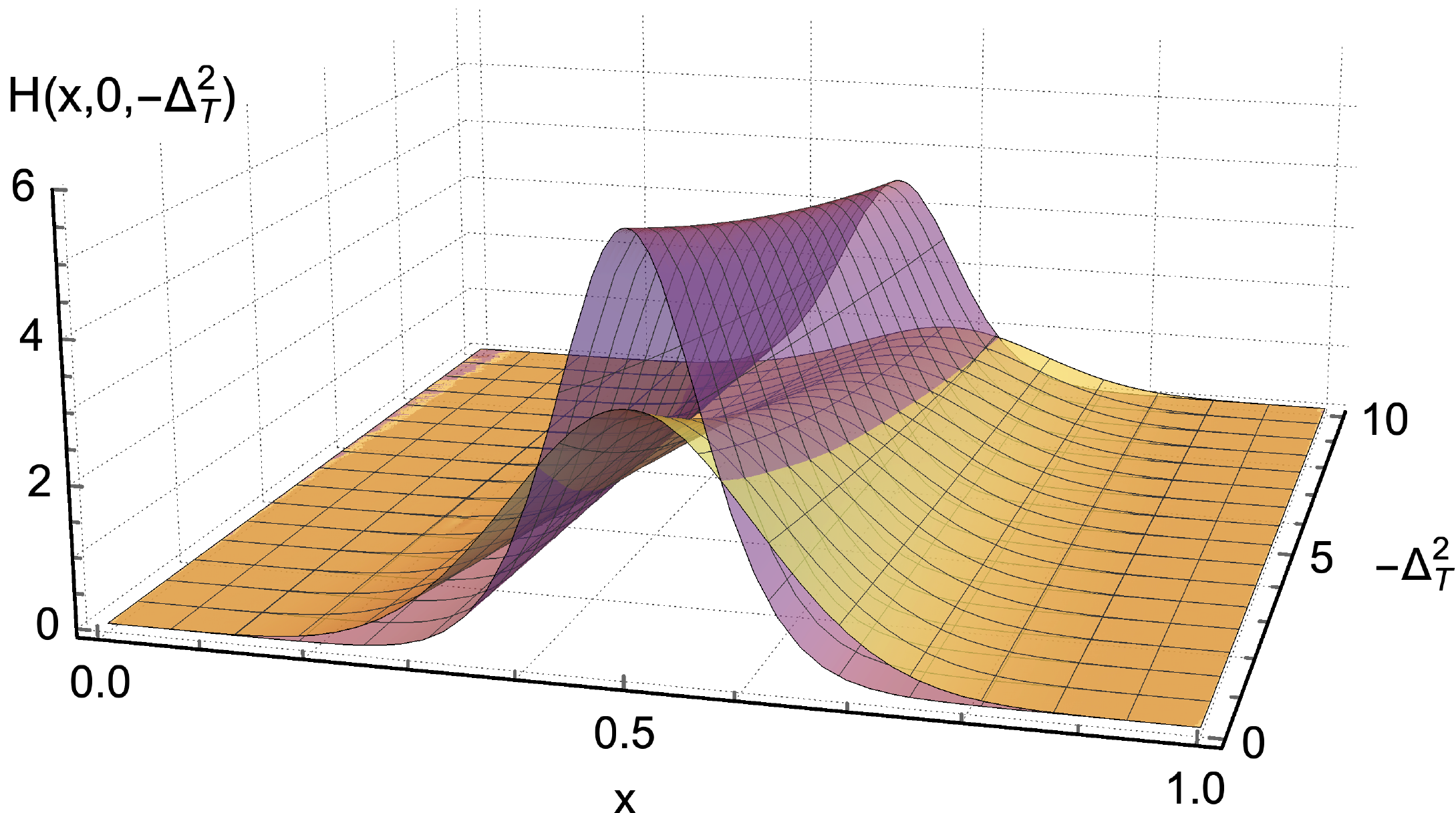} \\
\caption{The unpolarized GPD $H_M^q(x,\xi=0,t)$ of $\eta_c$ (yellow surface lower at $x=0.5$) and $\eta_b$ (purple surface upper at $x=0.5$) at scales around their masses.}
\label{fig:H0}
\end{figure}

We plot  $H_{\eta_c}^c(x,0,t;\mu_0)$ (yellow surface) and $H_{\eta_b}^b(x,0,t;\mu_0)$ (purple surface)  in Fig.~\ref{fig:H0}. Here we rescale the LF-LFWFs of $\eta_c$ and $\eta_b$ so that the PDF, which relates to the GPD as $f^q_M(x;\mu_0)=H^q_M(x,0,0;\mu_0)$, is normalized to unity. This approximation is based on the finding that higher Fock-states contribute little to the normalization, as shown in Table.~\ref{tab:N}.  However, the value of $\mu_0$ is priorly unknown. Phenomenologically, it is usually determined by comparing the first moment of PDF with experiment or lattice QCD \cite{Shi:2018mcb,Bednar:2018mtf}, but this is infeasible in the heavy sector due to the lack of experiment data. In this paper, we follow \cite{Lan:2019img} and presume that $\mu_0$ to be around $2 m_q$. That corresponds to $\mu_0 \approx 2.6$ GeV for $\eta_c$ and  $\mu_0 \approx 8.6$ GeV for $\eta_b$, close to the natural $\eta_c$ and $\eta_b$ mass scales. We remind that the various parton distributions studied below are all implicitly associated with such a scale, which will not be written  explicitly. %\footnote{The results of Table.~\ref{tab:N} is also associated with this hadronic scale $\mu_0$. For pion, $\mu_0\approx 2 m^{\rm Dr}_{u/d}=800$ MeV. The $m^{\rm Dr}_{u/d}$ is the dressed quark mass satisfying $M_{u/d}[(m^{\rm Dr}_{u/d})^2]=m^{\rm Dr}_{u/d}$  with $M_{u/d}$ the u/d quark mass function.}

%===============================================================================

\begin{figure}[tbp]
\centering\includegraphics[width=0.98\columnwidth]{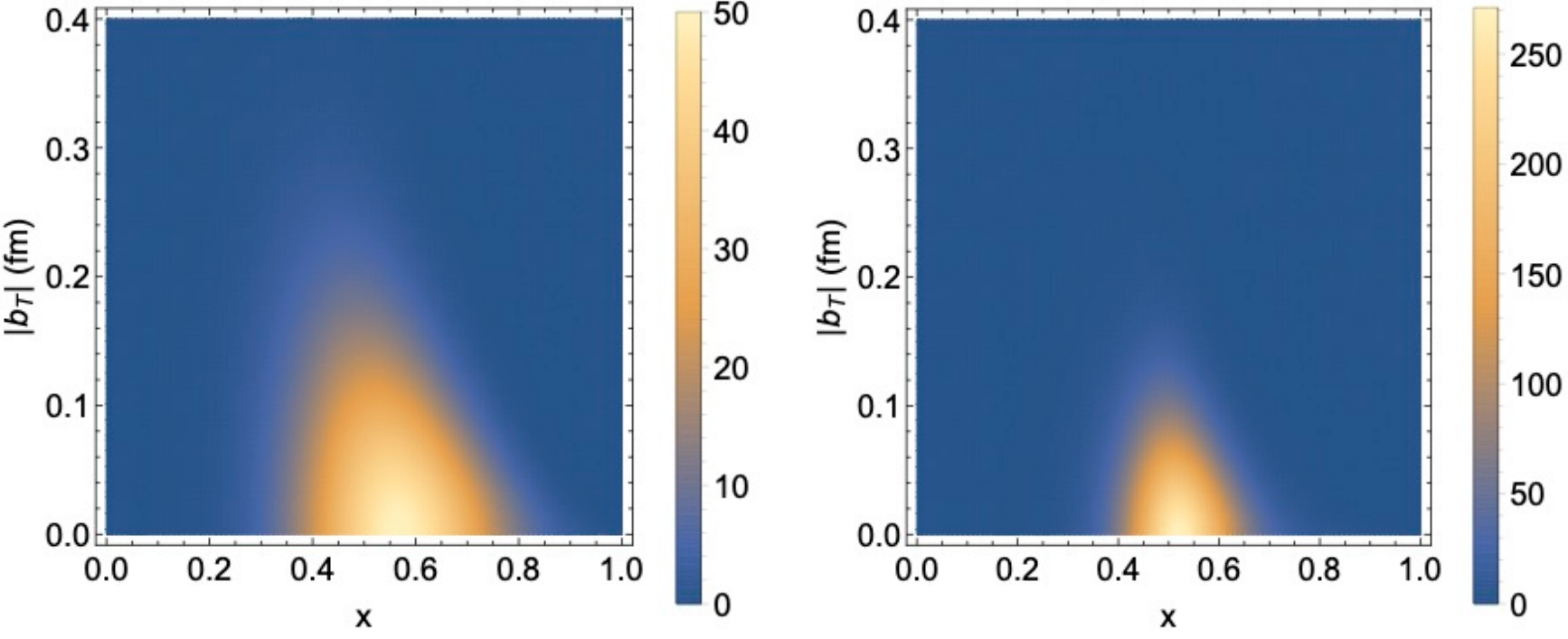} \\
\caption{\looseness=-1
The impact parameter dependent GPD $\rho^q_M(x,\vect{b}_T^2)$ of $\eta_c$  and $\eta_b$. The $\rho^c_{\eta_c}(x,\vect{b}_T^2)$ is on the left column and $\rho^b_{\eta_b}(x,\vect{b}_T^2)$ on the right.
}
\label{fig:rho}
\end{figure}

\begin{figure}[tbp]
\centering\includegraphics[width=0.94\columnwidth]{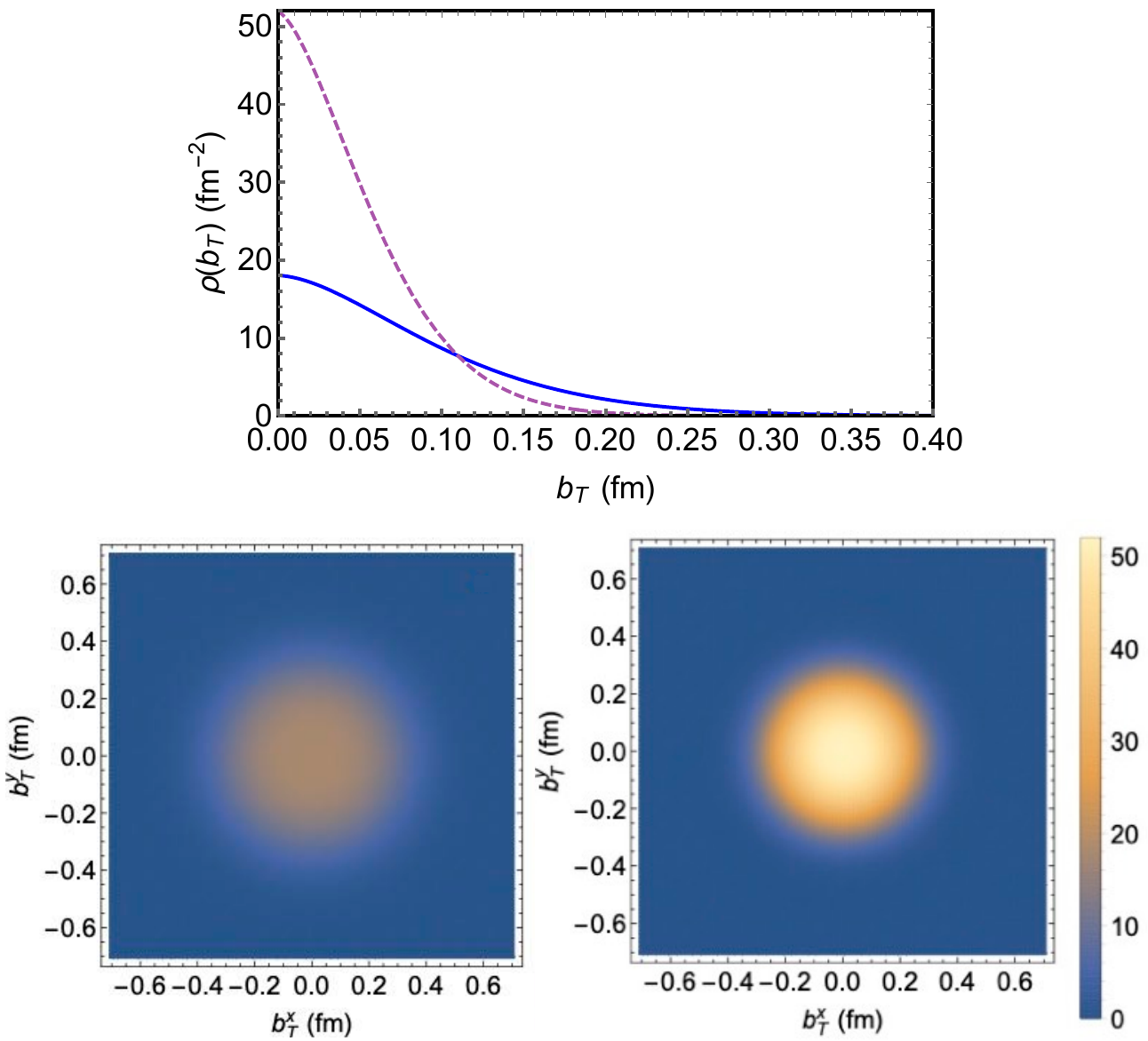} \\
\caption{\looseness=-1 \emph{Upper panel:} The transverse spatial distribution of charm quark within $\eta_c$ (blue solid) and bottom quark within $\eta_b$ (purple dashed) on the light front. \emph{Lower panel:} The density plots of the spatial distribution, with $\eta_c$ on the left and $\eta_b$ on the right.}
\label{fig:rho0}
\end{figure}

The two-dimensional Fourier transform of $H_M^q(x,0,\Delta_T^2)$ gives the IPD GPD
\begin{align}
\rho_M^q(x,\vect{b}_T^2) = 
\int \frac{d^2 \vect{\Delta}_T}{(2\pi)^2}\,H_M^q(x,0,-\vect{\Delta^2_T})\,e^{i \vect{b}_T \cdot \vect{\Delta}_T }, \label{eq:ipdfourier}
\end{align}
which characterizes the probability density of partons on the transverse plane with  the light-cone momentum fraction $x$ and the impact parameter $\vect{b}_T$ \cite{Burkardt:2002hr}. The $\vect{b}_T$ is defined as the separation between the active parton and the origin of transverse center of momentum, i.e., $\vect{b}_{T} \equiv \vect{b}_{T,1}=\vect{r}_{T,1}-\vect{R}_T$. In a two-body picture, $\vect{R}_T = x\,\vect{r}_{T,1} + (1-x)\,\vect{r}_{T,2}$, with $\vect{r}_{T,i}$ the transverse position of $i$-th parton. We show our results in Fig.~\ref{fig:rho}, with the left column devoted to charm quark in $\eta_c$, and the right column for bottom quark in $\eta_b$. The charm quark in $\eta_c$ is more broadly distributed in both $x$ and impact parameter $\vect{b}_T$ as compared to bottom quark in $\eta_b$ at their hadronic scales. We then integrate the $x-$dependence in the IPDs and obtain their spatial parton distribution $\rho_M^{q,(0)}(\vect{b_T})=\int_0^1 dx \rho_M^q(x,\vect{b_T}^2)$. The result is plotted in the upper panel of Fig.~\ref{fig:rho0} and the lower panel display their density plots. We determine their mean squared impact parameters $\langle \vect{b}_T^2 \rangle^q= \int d \vect{b}_T^2 \vect{b}_T^2 \rho^{q,(0)}(\vect{b}_T^2)$ to be $\langle \vect{b}_T^2 \rangle^c_{\eta_c}=(0.157\ {\rm fm})^2$ and $\langle \vect{b}_T^2 \rangle^b_{\eta_b}=(0.092\ {\rm fm})^2$. Comparing with our results on pion and kaon $\langle \vect{b}_T^2 \rangle^u_{\pi}=(0.332\ {\rm fm})^2$, $\langle \vect{b}_T^2 \rangle^u_{K}=(0.361\ {\rm fm})^2$ and $\langle \vect{b}_T^2 \rangle^s_{K}=(0.283\ {\rm fm})^2$ from \cite{Shi:2020pqe}, one finds the heavy mesons are considerably more compact in the transverse plane on the light front. 

By definition, the $\langle \vect{b}_T^2 \rangle^q_M$ can be regarded as the square of quark distribution radius in the light-cone frame. Within the leading Fock-state truncation and for a charged pseudoscalar meson, for instance the $\pi^+$, the light-cone charge radius of $\pi^+$ is $\langle r^2_{c,LC} \rangle_{\pi^+}=e_u \langle \vect{b}_T^2 \rangle^u_{\pi^+}+e_{\bar{d}} \langle \vect{b}_T^2 \rangle^{\bar{d}}_{\pi^+}=\langle \vect{b}_T^2 \rangle^u_{\pi^+}$. It is related to the $\pi^+$ charge radius in the Breit frame as $\langle r^2_{c,Br} \rangle_{\pi^+}=\frac{3}{2}\langle r^2_{c,LC} \rangle_{\pi^+}$. In the heavy quarkonium, the charge radius vanishes for charge conjugation. So we can follow \cite{Maris:2006ea,Li:2017mlw} and define a fictitious charge radius for $\eta_c$ in the Breit frame as $\langle r^2_{c,Br} \rangle_{\eta_c}=\frac{3}{2}\langle \vect{b}_T^2 \rangle^c_{\eta_c}=(0.192$ fm$)^2$. Note that the standard DS-BSEs computation  yields $\langle r^2_{c,Br} \rangle_{\eta_c}=(0.219$ fm$)^2$ \cite{Maris:2006ea}, in which infinitely many dressing and loop Feynman diagrams are calculated thus higher Fock-states are incorporated. In this sense, the deviation between $0.192$ fm and $0.219$ fm can be resorted to the higher Fock-states. Nevertheless, the deviation is less than 15\%, suggesting roughly the error brought by leading Fock-state truncation in $\eta_c$. We anticipate the deviation would be even smaller for heavier $\eta_b$.

%===============================================================================

\begin{figure}[tbp]
\centering\includegraphics[width=\columnwidth]{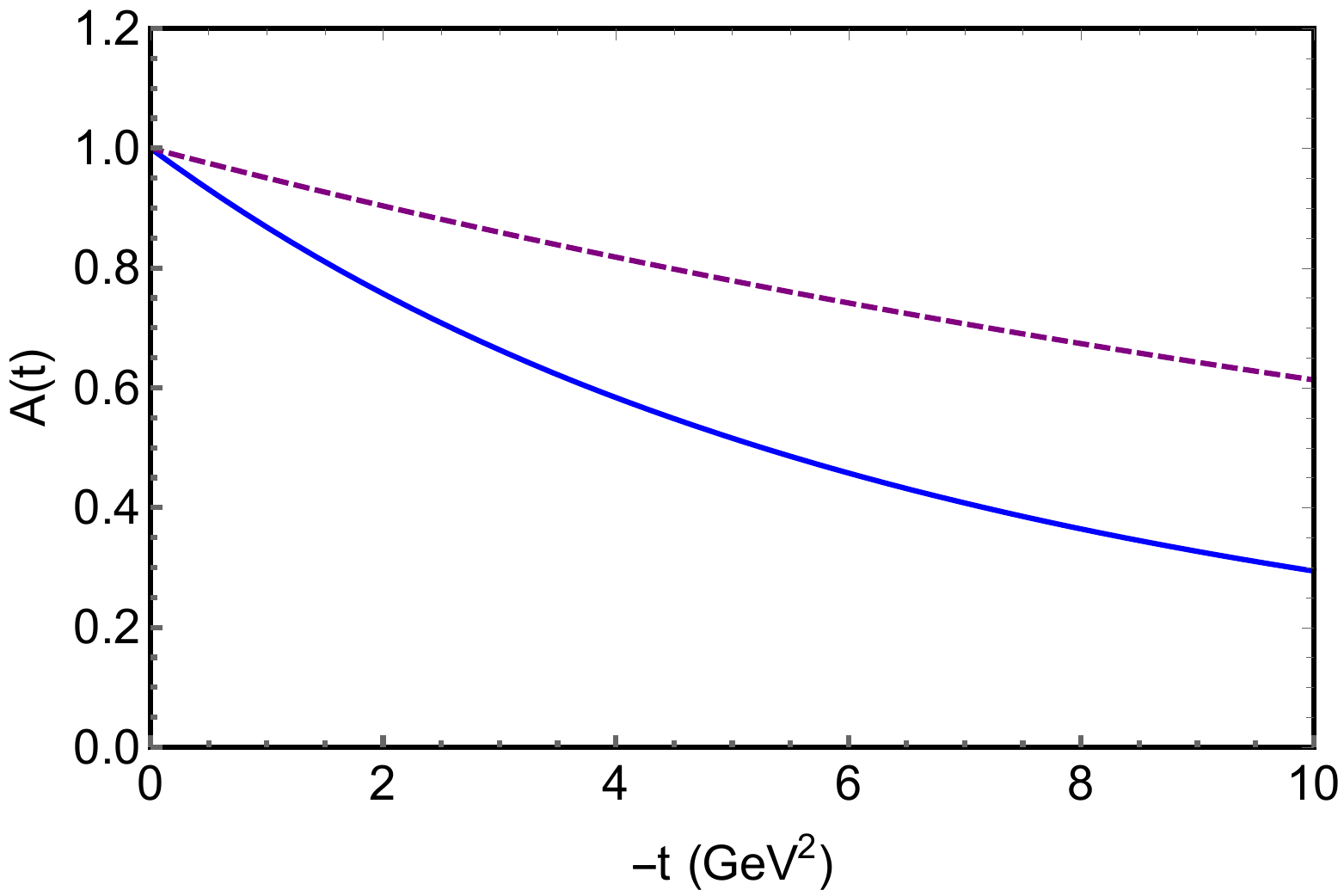} 
\caption{\looseness=-1 The gravitational form factor $A(t)$ (defined in Eq.~(\ref{eq:EMT}) of $\eta_c$ (blue solid) and $\eta_b$ (purple dotted) in the leading Fock-state truncation.
\label{fig:GFF}}
\end{figure}

The gravitational form factor of $\eta_c$ and $\eta_b$ can be connected with the $x-$moment of the GPD at $\xi=0$, i.e.,
\begin{align}
\int_{-1}^1dx\, x\, H^q(x,0,t) &= A^q_{2,0}(t).
\label{eq:theta2}
\end{align}
The $A_{2,0}^q(t)$ denotes the quark's contribution to the hadron's gravitational form factor $A(t)$, which enters the general decomposition of the matrix element of energy-momentum tensor (EMT) of spin-0 states~\cite{Donoghue:1991qv,Polyakov:2018zvc,Freese:2021mzg}
\begin{align}
\langle M(p')|T^{\mu\nu}(0)|M(p)\rangle&=\frac{1}{2}[P^\mu P^\nu A(t) \nonumber \\
&\hs*{8mm}
+(g^{\mu \nu}q^2-q^\mu q^\nu)C(t)].
\label{eq:EMT}
\end{align}
with $P=p+p'$, $q=p'-p$ and $t=q^2$. The $A(t)$ is related to the $x-$moment of GPDs by $A(t) = \sum_a A_{2,0}^a(t;\mu)$, where the $a=q,g$ runs through all partons. While the individual parton contributions $A_{2,0}^a(t;\mu)$ are scale dependent, their summation $A(t)$ is not. In our case, as we've approximated the $\eta_c$ and $\eta_b$ solely with the $q\bar{q}$ Fock-state, the gluon contribution $ A_{2,0}^g(t;\mu_0)$ vanishes  at initial scale $\mu_0$, thus $A(t)=2 A_{2,0}^q(t;\mu_0)$. We show our result for $\eta_c$ and $\eta_b$ in Fig.~\ref{fig:GFF}. 

%===============================================================================
%===============================================================================

The light-cone energy radius (or mass radius) $\lf< r_{E,{\rm LC}}^2 \rg>$ can be defined as the mean value of $\vect{r}_\perp^2$ weighted by EMT in the light-cone frame, namely $\frac{1}{2 P^+} \langle T^{++}\rangle_{\textrm{LC}}(\vect{r}_\perp)$,  where 
\begin{align}
\langle T_{\mu \nu} \rangle_{\textrm{LC}}(\vect{r}_\perp)=\int \frac{d^2 \Delta_\perp}{2 P^+ (2 \pi)^2} e^{-i \vect{\Delta}_\perp  \vect{r}_\perp}\langle p'|T_{\mu\nu}(0)|p\rangle.
\end{align}
It is thus related to the gravitational form factor $A(t)$ by~\cite{Freese:2019bhb} 
\begin{align}
\lf< r_{E,{\rm LC}}^2 \rg> = -4\,\lf.\frac{\partial\, A(Q^2)}{\partial Q^2} \rg|_{Q^2=0},
\end{align}
with $Q^2=-t$. For $\eta_c$ and $\eta_b$, we find the values to be $\langle r_{E,{\rm LC}}^2\rangle_{\eta_c} =(0.150$ fm)$^2$ and $\langle r_{E,{\rm LC}}^2 \rangle_{\eta_b} =(0.089$ fm)$^2$ respectively. It's interesting to compare them with the quark distribution radius $ \langle \vect{b}_T^2 \rangle^c_{\eta_c}=(0.157$ {\rm fm})$^2$ and $\langle \vect{b}_T^2 \rangle^b_{\eta_b}=(0.092$ {\rm fm})$^2$ above. The two radii are very close, with the light-cone energy radius a bit smaller. Such behaviors are in agreement with the finding of \cite{Li:2017mlw} in heavy mesons.

\begin{figure}[htbp]
\centering\includegraphics[width=\columnwidth]{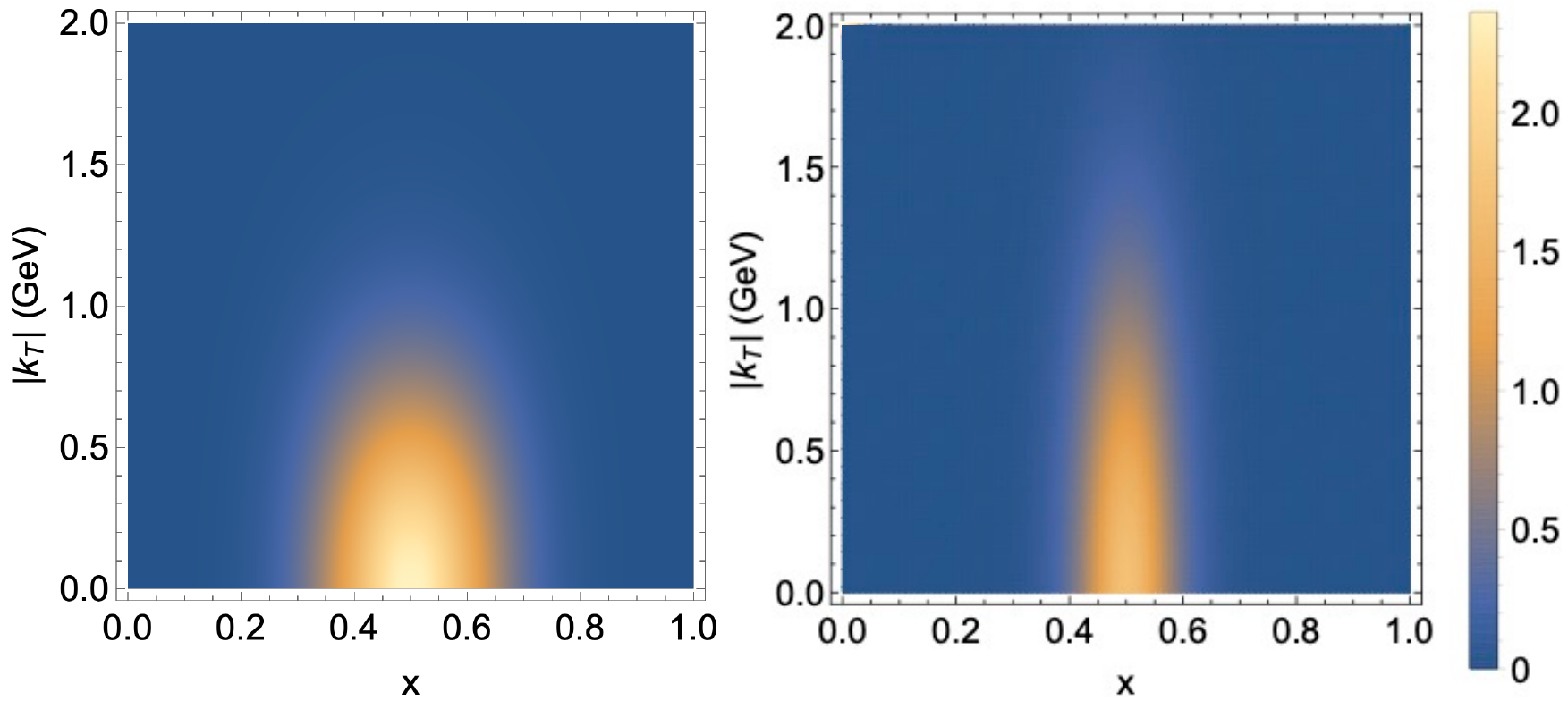} \\
\caption{
Density plots of the unpolarized TMD PDF $f_1^q(x,\vect{k}_T^2)$ of $\eta_c$ (left) and $\eta_b$ (right).}
\label{fig:TMD}
\end{figure}

We finally investigate the transverse momentum distribution of quarks within $\eta_c$ and $\eta_b$. The unpolarized leading-twist TMD of pseudoscalar meson is defined as
\begin{align} 
f_{1}(x,\vect{k}_T^2)&=\int\frac{d \xi^-d^2\vect{\xi}_T}{(2 \pi)^3}\ e^{i(\xi^-k^+-\vect{\xi}_T\cdot \vect{k}_T)}\nonumber\\
&\hspace{28mm}
\langle P|\bar{\psi}(0)\gamma^+\psi(\xi^-,\vect{\xi}_T)|P\rangle,
\end{align}
with the gauge link omitted. At hadronic scale, its overlap representation reads~\cite{Pasquini:2014ppa}
\begin{align}
f^q_1(x,\vect{k}_T^2) = \frac{1}{(2 \pi)^3} 
\left[\lf|\psi_0(x,\vect{k}_T^2)\rg|^2 + \vect{k}_T^2 \lf|\psi_1(x,\vect{k}_T^2)\rg|^2\right].
\label{eq:tmd}
\end{align}
 We show in Fig.~\ref{fig:TMD} their density plots. The result of $\eta_c$ is on the left while that of  $\eta_b$ on the right. These TMDs share the characteristics of the LF-LFWFs,  i.e., they peek at the center $x=0.5$ and are narrowly distributed in $x$. Comparing between the $\eta_c$ and $\eta_b$, we find the heavier bottom quarks are more centered around  $x=0.5$ while more broadly distributed in $\vect{k}_T$. Their mean transverse momentum of the valence quarks $\langle k_T \rangle=\int dx d^2 \vect{k}_T f_1^q(x,\vect{k}_T^2) |\vect{k}_T|$ is $\langle  k_T  \rangle_{\eta_c}=0.65$ GeV and $\langle  k_T  \rangle_{\eta_b}=1.02$ GeV. 

\begin{figure}[htbp]
\centering\includegraphics[width=\columnwidth]{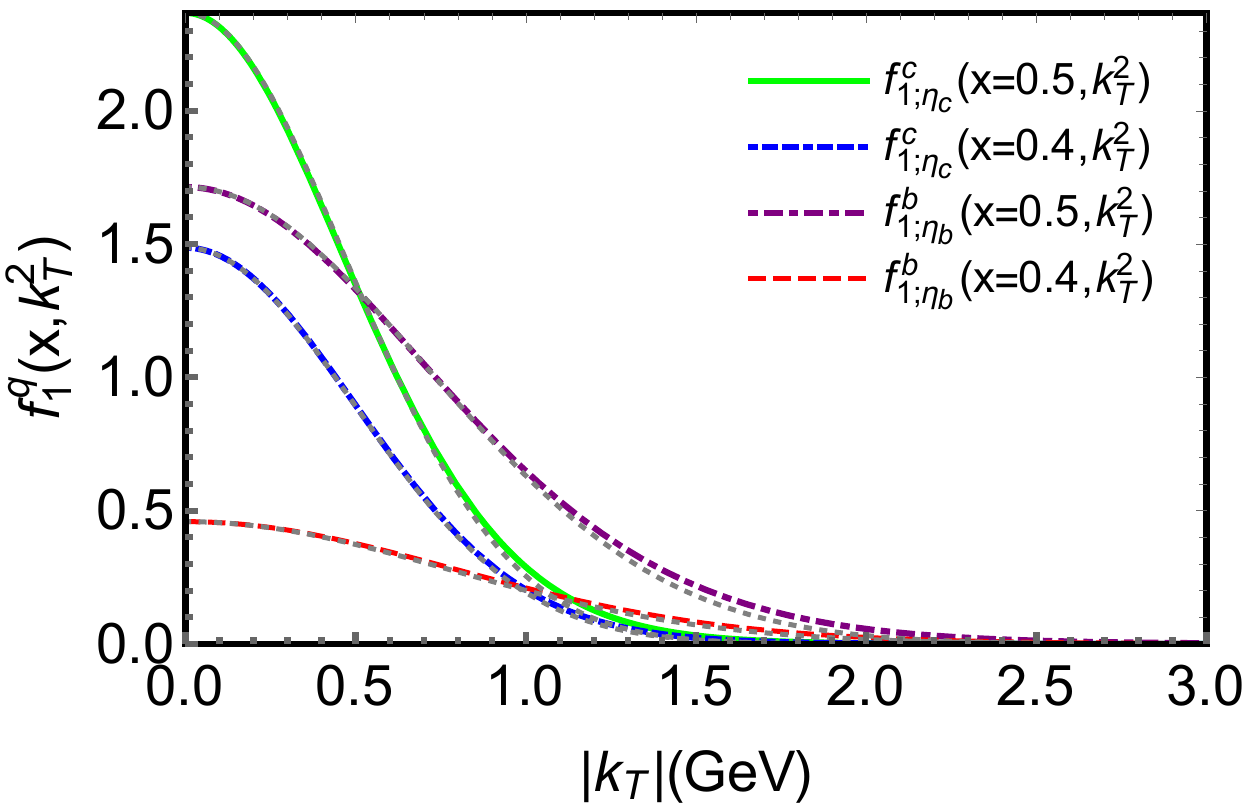} \\
\caption{
The $|\vect{k}_T|$-dependence of $\eta_c$'s and $\eta_b$'s unpolarized TMD at different $x$ values. The line styles are indicated in the plot. The accompanying  gray dotted curves are fitting curves in Gaussian form.}
\label{fig:gaussian}
\end{figure}

It's also interesting to look into the form of the  transverse momentum dependence within the TMD, which we demonstrate with Fig.~\ref{fig:gaussian}. For the past years, the Gaussian and/or Gaussian-based $|\vect{k}_T|$-dependent models have been very popular in parameterizing the TMDs of pion and nucleon in the light quark sector \cite{DAlesio:2004eso,Anselmino:2005nn,Collins:2005ie,Schweitzer:2010tt,Aybat:2011zv,Wang:2017zym,Bacchetta:2017gcc,Shi:2018zqd}. Here we explore its validity in the heavy sector with the Gaussian form $f_{\textrm G}(x,\vect{k}_T^2)=N e^{-\vect{k}_T^2/\langle \vect{k}_T^2(x) \rangle}$. We find the fitting curves (gray dotted) largely overlap with the original curves. The fitting parameters were determined to be $\langle \vect{k}_T^2(x=0.5)\rangle=(0.67$ GeV)$^2$ and $\langle \vect{k}_T^2(x=0.4)\rangle=(0.7$ GeV)$^2$ for $\eta_c$, and  $\langle \vect{k}_T^2(x=0.5)\rangle=(1.0$ GeV)$^2$ and $\langle \vect{k}_T^2(x=0.4)\rangle=(1.1$ GeV)$^2$ for $\eta_b$ respectively. These values are close to the total mean transverse momentum $\langle k_T \rangle_{\eta_c}=0.65$ GeV and $\langle k_T \rangle_{\eta_b}=1.02$ GeV obtained above, suggesting weak $x-$dependence in fitting parameter $\langle \vect{k}_T^2(x) \rangle$.

%===============================================================================

\section{CONCLUSION\label{sec:con}}
We study the leading Fock-state light front wave functions of the light and heavy pseudoscalar mesons, using a unified framework of rainbow-ladder DS-BSEs and light front projection method. The LF-LFWFs of the pion at masses of $310$ MeV and $690$ MeV, and those of the heavy $\eta_c$ and $\eta_b$ are reported for the first time within DS-BSEs. 

The LF-LFWFs of pion at physical mass and $310$ MeV are broadly distributed in $x$ and close to each other, while at the mass of $690$ MeV which sits at the strange quark point, i.e., the fictitious $\pi_{s\bar{s}}$, the LF-LFWFs get distinctly narrower.  This trend continues in the heavy sector for the $\eta_c$ and $\eta_b$. Such property is reflected in the twist-2 PDA which is the $\vect{k}_T-$integrated LF-LFWF. General agreement is found between our pion PDAs at different masses and those from lattice QCD \cite{Zhang:2020gaj}. In the heavy sector, where lattice QCD is absent, our PDAs of $\eta_c$ and $\eta_b$ are comparable with those from sum rule prediction \cite{Zhong:2014fma}. 

The contribution of the LF-LFWFs to meson states on the light front is further analyzed. As listed in Table.~\ref{tab:N}, the LF-LFWFs of pion contribute only 30\% to the Fock-state normalization, while in $\eta_c$ and $\eta_b$ they contribute more than 90\%. This indicates the existence of considerable higher Fock-states in pion, and meanwhile the dominant role of LF-LFWFs in determining heavy mesons. In particular, the later strongly suggests the leading Fock-state truncation as a reasonable means in dealing with $\eta_c$ and $\eta_b$. 

We thus take the leading Fock-state truncation, and study the unpolarized quark GPD and TMD of $\eta_c$ and $\eta_b$ using the light front overlap representation given in Eq.~(\ref{eq:Hoverlap})  and Eq.~(\ref{eq:tmd}). We associate the resolution scale to the sum of consistuent valence (anti)quark masses, i.e., $2 m_q$, and study the the spatial distribution of valence quarks with IPD GPD. It is found that heavier quarks are spatially more centered in heavier mesons. The study on gravitational form factor also reveals that the heavier meson has smaller energy radius in the light-cone frame. In the transverse momentum space, we find the heavier quark is more broadly distributed in $\vect{k}_T$, but more centered around longitudinal momentum fraction $x=0.5$. We also find the unpolarized TMD PDF of $\eta_c$ and $\eta_b$ can be approximated with $x-$independent Gaussian functions, as an extension to rudimentary phenomenological parameterizations of hadron TMD in the light sector.

To conclude, this paper delivers comprehensive insights into the LF-LFWFs of pseudoscalar mesons with varying current quark mass, as well as valence quark imaging of $\eta_c$ and $\eta_b$  in both position and momentum space.

\begin{acknowledgments}
This work is supported by the National Natural Science Foundation of China (under Grant No. 11905104) and the Strategic Priority Research Program of Chinese Academy of Sciences (Grant NO. XDB34030301).
\end{acknowledgments}

\appendix
\section{Parameterization of $\vect{S(k)}$ and $\vect{\Gamma(k;P)}$}\label{sec:app}
We list the parameterization parameters of $\vect{S(k)}$ and $\vect{\Gamma(k;P)}$ from Eqs.~(\ref{eq:gammapara}-\ref{eq:rho}) in Table.~\ref{tab:parapion} and Table.~\ref{tab:paraetaQ}.

\begin{table}[htbp]
\caption{Representation parameters for pion with different masses $m_\pi=130$ MeV, $310$ MeV and $690$ MeV. \emph{Upper panel}: Eq.~(\ref{eq:spara}) -- the pair $(x,y)$ represents the complex number $x+ i y$.  \emph{Lower panel}: Parameters of Eqs.~(\ref{eq:gammapara},\ref{eq:fpara},\ref{eq:rho}). (Dimensioned quantities are given in GeV).
\label{tab:parapion}
}
\begin{center}
\begin{tabular*}%{lcccc}
{\hsize}
{
@{\extracolsep{0ptplus1fil}}
c@{\extracolsep{0ptplus1fil}}
c@{\extracolsep{0ptplus1fil}}
c@{\extracolsep{0ptplus1fil}}
c@{\extracolsep{0ptplus1fil}}
c@{\extracolsep{0ptplus1fil}}}\hline
  & $z_1$ & $m_1$  & $z_2$ & $m_2$ \\
$u_{\pi^{130}}$ &    $(0.31, 0.25)$ & $(0.52, 0.27)$ & $(0.11, 0.0025)$ & $(-0.81, 0.71)$ \\
$u_{\pi^{310}}$ &    $(0.32, 0.23)$ & $(0.57, 0.29)$ & $(0.11, 0.016)$ & $(-0.82, 0.73)$ \\
$u_{\pi^{690}}$ &    $(0.31, 0.26)$ & $(0.75, 0.37)$ & $(0.10, 1.89)$ & $(-0.89, 0.011)$ \\
\hline
\end{tabular*}

\begin{tabular*}%{llcccccccc}
{\hsize}
{
l@{\extracolsep{0ptplus1fil}}
l@{\extracolsep{0ptplus1fil}}
c@{\extracolsep{0ptplus1fil}}
c@{\extracolsep{0ptplus1fil}}
c@{\extracolsep{0ptplus1fil}}
c@{\extracolsep{0ptplus1fil}}
c@{\extracolsep{0ptplus1fil}}
c@{\extracolsep{0ptplus1fil}}
c@{\extracolsep{0ptplus1fil}}
c@{\extracolsep{0ptplus1fil}}}\hline
   & $U_1$ & $U_2$ & $U_3$ &$n_1$ &$n_2$ &$n_3$ & $\sigma^i_1$ & $\sigma^i_2$ & $\Lambda$ \\[0.7ex]\hline
 E$_{\pi^{130}}$ & $9.88$ & $-5.93$ & $0$ & $6$
    & $8$ & $-$&1.76 &0.97&1.7 \\
   F$_{\pi^{130}}$ & $5.14$ & $3.65$ & $0$ &$6$
    & $8$ & $-$&0.65&-2.0& 1.7   \\
 G$_{\pi^{130}}$ & $4.15$ & $-11.36$ & $0$ & $8$
    & $10$ & $-$&-0.30 &-0.37&1.4 \\
   H$_{\pi^{130}}$ & $1.28$ & $2.57$ & $0$ &$6$
    & $8$ & $-$&0.74&0.36& 1.8  \\\hline
E$_{\pi^{310}}$ & $10.44$ & $-8.14$ & $0$ & $6$
    & $8$ & $-$&1.48 &1.13&1.7 \\
   F$_{\pi^{310}}$ & $5.01$ & $2.61$ & $0$ &$6$
    & $8$ & $-$&-0.085&-2.22& 1.7   \\
 G$_{\pi^{310}}$ & $2.92$ & $-11.38$ & $0$ & $8$
    & $10$ & $-$&-0.43 &-0.55&1.45 \\
   H$_{\pi^{310}}$ & $1.04$ & $2.09$ & $0$ &$6$
    & $8$ & $-$&0.20&-0.011& 1.86  \\\hline
 E$_{\pi^{690}}$ & $11.97$ & $-0.99$ & $0$ & $6$
    & $8$ & $-$&1.38 &9.20&2.0 \\
   F$_{\pi^{690}}$ & $4.67$ & $3.95$ & $0$ &$6$
    & $8$ & $-$&0.58&-1.51& 2.0   \\
 G$_{\pi^{690}}$ & $1.13$ & $-5.99$ & $0$ & $8$
    & $10$ & $-$&-0.49 &-0.36&1.7 \\
   H$_{\pi^{690}}$ & $0.55$ & $-0.78$ & $0$ &$8$
    & $10$ & $-$&0.36&1.15& 1.6  \\\hline
\end{tabular*}
\end{center}
\end{table}

\begin{table}[htbp]
\caption{Representation parameters for $\eta_c$ and $\eta_b$ mesons. \emph{Upper panel}: Eq.~(\ref{eq:spara}) -- the pair $(x,y)$ represents the complex number $x+ i y$.  \emph{Lower panel}: Parameters of Eqs.~(\ref{eq:gammapara},\ref{eq:fpara},\ref{eq:rho}).    (Dimensioned quantities are given in GeV).
\label{tab:paraetaQ}
}
\begin{center}
\begin{tabular*}%{lcccc}
{\hsize}
{
@{\extracolsep{0ptplus1fil}}
c@{\extracolsep{0ptplus1fil}}
c@{\extracolsep{0ptplus1fil}}
c@{\extracolsep{0ptplus1fil}}
c@{\extracolsep{0ptplus1fil}}
c@{\extracolsep{0ptplus1fil}}}\hline
  & $z_1$ & $m_1$  & $z_2$ & $m_2$ \\
$u$ &    $(0.47, 0.70)$ & $(1.84, 0.54)$ & $(0.018, 0.033)$ & $(-1.91, -1.16)$ \\
$b$ &    $(0.47, 0.66)$ & $(5.1, 0.74)$ & ----- & ----- \\
\hline
\end{tabular*}

\begin{tabular*}%{llcccccccc}
{\hsize}
{
l@{\extracolsep{0ptplus1fil}}
l@{\extracolsep{0ptplus1fil}}
c@{\extracolsep{0ptplus1fil}}
c@{\extracolsep{0ptplus1fil}}
c@{\extracolsep{0ptplus1fil}}
c@{\extracolsep{0ptplus1fil}}
c@{\extracolsep{0ptplus1fil}}
c@{\extracolsep{0ptplus1fil}}
c@{\extracolsep{0ptplus1fil}}
c@{\extracolsep{0ptplus1fil}}}\hline
 & $U_1$ & $U_2$ & $U_3$ &$n_1$ &$n_2$ &$n_3$ & $\sigma^i_1$ & $\sigma^i_2$ & $\Lambda$ \\[0.7ex]\hline
 E$_{\eta_c}$ & $7.14$ & $-8.07$ & $0.18$ & $5$
    & $6$ & $1$&-1.09 &-0.91&2.4 \\
   F$_{\eta_c}$ & $1.07$ & $0.21$ & $0.01$ &$5$
    & $6$ & $1$&-1.27&-3.86& 2.4   \\
 G$_{\eta_c}$ & $0.14$ & $-0.57$ & $0.0071$ & $5$
    & $6$ & $2$&-0.87 &-0.58&2.0 \\
   H$_{\eta_c}$ & $0.071$ & $-0.079$ & $0.0021$ &$5$
    & $6$ & $2$&-0.69&0.15& 1.9  \\\hline
 E$_{\eta_b}$ & $10.91$ & $-7.31$ & $0.44$ & $5$
    & $6$ & $1$&-1.51 &-0.74&3.6\\
   F$_{\eta_b}$ & $0.55$ & $-0.29$ & $0.013$ &$5$
    & $6$ & $1$&-1.89&-1.40& 3.4   \\
 G$_{\eta_b}$ & $0.0011$ & $-0.15$ & $0.0022$ & $5$
    & $6$ & $2$&-1.75 &-1.72&3.3 \\
   H$_{\eta_b}$ & $0.0076$ & $-0.0015$ & $0.00055$ &$5$
    & $8$ & $2$&-1.41&2.63& 3.4  \\\hline
\end{tabular*}
\end{center}
\end{table}

\clearpage
%===============================================================================

\bibliography{QQP}

\end{document}